# Breaking the Balance:
# Asymmetric Negative Voting in the 2020 Presidential Election

Bang Quan Zheng

University of Texas at Austin

2025

Abstract

While negative voting has traditionally been symmetrical across partisan lines, recent elections suggest an emerging asymmetry. This study builds on theories of affective polarization and retrospective voting to argue that negative voting is increasingly shaped by partisan differences in candidate evaluations and responsiveness to political crises. Using 2016 and 2020 Nationscape data, we show that while negative voting was relatively balanced in 2016—Clinton and Trump faced comparable levels of hostility—by 2020, Democrats and Independents exhibited significantly higher levels of negative voting against Trump, whereas Republicans' opposition to Biden was notably weaker than their hostility toward Clinton in 2016. This shift reflects partisan asymmetries in retrospective evaluations, as dissatisfaction with Trump's handling of the economy, race relations, and the COVID-19 pandemic disproportionately mobilized younger, educated Democrats and Independents. These findings challenge symmetric partisan negativity and show how crises amplify asymmetries in negative voting, reshaping elections.

[Word count: 6,867]

## 1 Introduction

Scholars have extensively documented rising polarization in the American electorate (Hetherington, 2001; McCarty et al., 2016; Suhay, 2015). At the core of this divide is affective polarization, where Democrats and Republicans increasingly view each other with hostility (Druckman et al., 2018; Iyengar et al., 2012; Robison & Moskowitz, 2019; Tyler & Iyengar, 2023). This animosity has reinforced partisan identity alignment, deepening ideological and social divisions (Bekafigo et al., 2019; Robison & Moskowitz, 2019; Weber & Klar, 2019). One key electoral manifestation of affective polarization is negative voting—casting a ballot primarily to oppose the out-party candidate rather than to support one's own (Fiorina & Shepsle, 1989). Traditionally, negative voting has been symmetric, with both parties exhibiting comparable levels of hostility toward opposing candidates. However, the 2020 presidential election marked a significant shift. Anti-Trump sentiment became the dominant force behind negative voting, particularly among younger, educated Democrats and Independents, whereas Republicans exhibited significantly less hostility toward Biden compared to Clinton in 2016. While negative partisanship has often been treated as a stable and symmetric feature of partisan competition, this study reveals that under certain conditions—such as extreme elite polarization and the presence of a uniquely polarizing incumbent—negative voting can become highly asymmetric. This suggests that rather than being an inherent feature of contemporary elections, negative partisanship may be contingent on contextual and candidate-specific factors, challenging existing assumptions about its long-term dynamics.

Donald Trump's presidency deepened partisan divisions, particularly through his policies on immigration, race, and his combative relationship with the media (Bekafigo et al., 2019). From a retrospective voting perspective, voters often rely on past performance to guide their choices, with negative perceptions exerting greater influence than positive evaluations. Negative partisanship further amplifies this effect: voters are more likely to cast a ballot against a disliked candidate than in favor of a preferred one (Garzia & Ferreira da Silva, 2022). Moreover, group polarization processes push attitudes to extremes as individuals engage in discussions within ideologically homogeneous networks (Robison & Moskowitz, 2019). These dynamics were especially pronounced in 2020, when dissatisfaction with Trump's handling of the economy, race relations, the COVID-19 pandemic, and his leadership style disproportionately mobilized younger, educated Democrats and Independents against him. In contrast, Republicans—particularly those in predominantly white, lower-income communities with less formal education—were less likely to engage in negative voting against Biden (Canipe et al., 2020; Pew Research Center, 2020a). While Biden ultimately secured 81.3 million votes—over 7 million more than Trump—much of this support stemmed from anti-Trump sentiment rather than strong enthusiasm for Biden himself (Jacobson, 2021). Understanding the conditions under which negative voting becomes asymmetric has important implications for the stability of party coalitions, voter mobilization strategies, and the long-term trajectory of affective polarization in American politics.

This study contributes to the literature in three key ways. First, it highlights the growing asymmetry in negative voting, challenging the prevailing assumption of balanced hostility toward out-party candidates. Second, it demonstrates how anti-Trump sentiment uniquely shaped electoral polarization, particularly among younger and more educated voters. Finally, it provides empirical evidence on the evolution of negative voting under Trump's presidency, illustrating how retrospective evaluations and demographic divides influenced electoral outcomes amid a period of political crisis and uncertainty.

## 2 Controversy and Unpopularity Surrounding Trump

The 2020 election was largely shaped by public reactions to Donald Trump, with opinions deeply polarized along party lines (Jacobson, 2021). Anti-Trump sentiment encompasses diverse attitudes toward Trump's performance as president, particularly regarding the COVID-19 pandemic, the economy, race relations, and his overall leadership. Trump remains one of the most divisive figures in U.S. history, playing a central role in exacerbating political polarization (Jacobson, 2021).

### *2.1 Race Relations*

Trump's rhetoric and policies on immigration, especially his hardline stance on U.S.-Mexico border security, drew significant backlash. His disparaging comments about Mexican immigrants and calls for mass deportations alienated many in the Latino community (Bekafigo et al., 2019). This inappropriate rhetoric and controversial policies sparked strong opposition, further intensifying the already polarized attitudes toward immigration. In addition, in January 2017, shortly after taking office, Trump signed an executive order that temporarily banned the entry of 218 million people from seven predominantly Muslim countries: Iran, Iraq, Libya, Somalia, Sudan, Syria, and Yemen (Diamond, 2017). The initial travel ban sparked widespread protests, legal challenges, and debates about its constitutionality. The travel ban faced criticism for targeting Muslim-majority countries and was perceived by many as discriminatory. The controversy surrounding the ban continued as it underwent several iterations and faced legal battles, with opponents arguing that it violated the principles of religious freedom and equal protection under the law. Trump's divisive rhetoric appealed to some conservatives and his supporters who tend to perceive the world in a similar manner, emphasizing an "us-versus-them" mentality (MacWilliams, 2016). By 2018, Trump's unpopularity was evident in the midterm elections, where congressional Democrats achieved a 9-point advantage over Republicans in the nationwide popular vote for the U.S. House of Representatives (Igielnik et al., 2021).

### *2.2 The Outbreak of Pandemic*

The COVID-19 pandemic had infected over 4.3 million Americans and claimed approximately 37,000 lives by November 2020 (Beer, 2020). Amid the pandemic, Trump and his administration faced widespread criticism for their response, which was marked by significant mismanagement (Abutaleb et al., 2020). His personal approach to the crisis was often seen as unsettling, reflected in record-high turnover among senior staff and Cabinet members. Trump's approach to the pandemic was marked by a consistent pattern of misinformation, indifference, and a reluctance to take accountability (Jacobson, 2020; Jacobson, 2021). He was notably hesitant to advocate for mask-wearing during the crisis, which many perceived as a crucial public health measure. For some, his actions suggested he had effectively abandoned efforts to control the pandemic (Hart, 2022). As a result, Trump received lower ratings for providing accurate information about the pandemic to the public. According to Pew research survey conducted in April 2020, only 57 percent of respondents believed he was doing only a fair or poor job, with about 41 percent rating his performance as poor. Conversely, around 21 percent of respondents said he did an excellent or good job in delivering accurate information (Pew Research Center, 2020b). Opinions on Trump's handling of the pandemic are divided across different levels of educational attainment: 71 percent of those with a bachelor's degree or more education said Trump was too slow to take action, while 62 percent of those without a college degree said the same (Pew Research Center, 2020b). Trump's mishandling of the pandemic caused a backlash among minority voters. While he gained support from Black voters and Latinos in 2016—particularly in Texas and Florida, where their votes contributed to his victories—both states later experienced severe impacts from the COVID-19 pandemic, including high infection and death rates (Canipe et al., 2020).

### *2.3 Sudden Economic Downturn*

The pandemic-induced recession brought severe economic hardship, with over ten million Americans unemployed, many contended that Trump could have exerted pressure on Congressional Republicans to expedite economic relief efforts. However, he chose not to take such action. A majority of respondents expressed that Trump performed only a fair or poor job in addressing key aspects of the pandemic response: 54 percent criticized his handling of the economic needs of ordinary Americans affected by the outbreak, 54 percent disapproved of his collaboration with state governors, and 55 percent were dissatisfied with his response to the needs of hospitals, doctors, and nurses (Pew Research Center, 2020b). Consequently, Trump's presidency heightened the animosity between the Democratic and Republican parties, showcasing increased levels of polarization. One of the defining political trends during Trump's tenure was the exodus of independents, suburbanites, white women, and college-educated voters from the Republican Party (Bennett & Rogers, 2020). As noted by Garzia and Ferreira da Silva (2022) in their research, 61 percent of negative voters reported that the U.S. economy had worsened or significantly worsened compared to the previous year. In this challenging economic environment, Trump, as an incumbent seeking re-election in 2020, faced a significant retrospective voting disadvantage, further exacerbated by negative voting. This can largely be attributed to Trump's campaign strategy in 2020, similar to his approach in 2016. As noted by Hart (2022) Trump maintained his *paranoid style* in his campaign, ignoring the reality that thousands of Americans were dying from COVID-19 every day and many more were losing their jobs (see Hart, 2022, p. 13). In contrast, Biden secured his victory by broadening his party's appeal among suburban voters, middle- and upper-income communities, and areas with a high percentage of college graduates (Canipe et al., 2020). In the 2020 election, therefore, the combination of high turnout and shifting voter preferences was enough for Biden to narrowly flip a handful of key battleground states—Pennsylvania, Michigan, and Wisconsin—that had tipped in Trump's favor in 2016.

# 3 Negative Voting & Affective Polarization

For many political scientists, Donald Trump became a polarizing figure, deepening divisions between liberals and conservatives, as well as between Democrats and Republicans. Three major theoretical perspectives help explain negative voting and its impact on affective polarization. The first perspective focuses on rational choice and retrospective voting. According to this view, voters assess the incumbent's past performance when making electoral decisions, with negative evaluations often exerting a stronger influence than positive ones (Fiorina & Shepsle, 1989; Kernell, 1977). For instance, economic downturns disproportionately harm in-party candidates in congressional elections (Bloom & Price, 1975). From a psychological perspective, Lau (1982, 1985) argued that negative voting is driven by loss aversion, where voters aim to prevent undesirable outcomes—such as the election of a disliked candidate—rather than actively support a preferred one. This defensive voting strategy reflects patterns observed in midterm losses for the president's party, often linked to turnout asymmetries (Kernell, 1977).

The second perspective, rooted in political psychology, highlights the role of cognitive dissonance in negative voting. *The American Voter* (1960) has pointed out half a century ago that "As long as public affairs go well, there is little to motivate the electorate to connects of the wider environment with the actors of politics, and the success of an administration are likely to go virtually unnoticed by the mass public. But when events of the wider environment arouse strong public concern, the electorate is motivated to connect them with the actors of politics—typically, with the incumbent party" (p. 556). Further expanding on this psychological concept, Gant and Davis (1984) applied cognitive dissonance theory to explain negative voting. They posited that rejecting a candidate, rather than expressing a positive preference for another, serves as a psychological mechanism to reduce cognitive dissonance. This dissonance may arise from conflicting feelings about the choices available or dissatisfaction with the political landscape as a whole. Voters align their actions with negative attitudes to minimize discomfort caused by dissatisfaction with political choices or the broader political environment.

The third framework focuses on affective polarization and negative partisanship, where negative voting is driven by a combination of in-party loyalty and out-party hostility. Recent research suggests that negative voting intensifies when aversion to the opposing candidate outweighs support for one's own candidate (Garzia & Ferreira da Silva, 2022). These dynamic underscores the strength of out-party hostility as a more powerful motivator for voting behavior than in-party allegiance. However, single snapshot data do not fully capture whether shifts in partisan identification reflect deeper changes in voters' attitudes between 2016 and 2020. Republicans dissatisfied with Trump's performance—especially his handling of the pandemic and race relations—may have voted against him without rejecting the Republican Party itself. Similarly, independents, who lack strong partisan loyalties, may have cast negative votes to express their anti-Trump sentiment.

Social group affective polarization further reinforces negative voting. Robison and Moskowitz (2019) found that partisans consistently rated in-party groups more favorably than out-party groups, with this gap widening over time. Evaluations of social groups correspond to increasingly polarized party thermometer ratings. Bekafigo et al. (2019) argue that group polarization drove more extreme attitudes toward Trump after exposure to like-minded discussions, intensifying the "echo chamber" effect. As Lau (1985) noted, negative information carries greater weight than positive information, particularly when it contradicts expectations. This dynamic may explain why younger and more educated voters are more likely to engage in negative voting, driven by heightened polarization and social group distinctions.

Building on these theories, we propose the following hypotheses:

***H1.*** Individuals who perceived economic decline were more likely to cast a negative vote against Trump in the 2020 election, as economic dissatisfaction fueled opposition to the incumbent.

***H2.*** Individuals with negative evaluations of Trump's leadership and policies were more inclined to engage in negative voting against him, reinforcing the centrality of candidate-specific assessments in electoral behavior.

***H3.*** Strong out-partisan animosity toward Republicans heightened the likelihood of casting a negative vote against Trump, reflecting the growing role of affective polarization in shaping voter decision-making.

***H4.*** Younger voters were more likely to engage in negative voting against Trump, as generational differences in political attitudes and issue priorities intensified opposition to his presidency.

***H5.*** Higher levels of education were associated with an increased propensity for negative voting against Trump, as more educated voters tended to express stronger opposition to his policies and leadership style.

## 4 Data & Method

The data for this study come from the Nationscape dataset (Democracy Fund Voter Study Group, 2021), which provides nationwide multi-wave data covering the 2016 and 2020 presidential elections. The first wave was collected in December 2016, with 8,000 respondents, of whom 4,943 were reinterviewed in November 2020. Additionally, the 2016-2020 dataset includes 4,483 new respondents, many of whom are first-time or young voters. Administered by YouGov, the data uses a matching procedure that aligns each observation with a similar one from YouGov's panel sample based on demographic characteristics.

### *4.1 Dependent Variables and Key Independent Variables*

The study focuses on four key dependent variables related to voting motivations. In the Nationscape data, respondents were asked about their motivations for voting in the 2016 and 2020 elections, specifically whether their vote was for or against a candidate. In 2016, respondents were asked whether their vote for Hillary Clinton or Donald Trump was primarily a vote in favor of their chosen candidate or against the opponent. In 2020, a similar question was asked regarding Joe Biden and Donald Trump.

Based on these questions, we created four binary dependent variables: Voting for Trump to oppose Clinton (2016), voting for Clinton to oppose Trump (2016), voting for Biden to oppose Trump (2020), voting for Trump to oppose Biden (2020). Each variable is binary, coded as 1 for the specified vote and 0 otherwise.

In addition, many studies assume stable partisanship, but some individuals may have shifted affiliations between 2016 and 2020 due to candidate or policy disapproval. To account for this, we analyze respondents from both survey waves. Among them, 91 percent of Democrats and 92 percent of Republicans retained their affiliation, while 6 percent shifted to independent status, and 6 percent of independents became Democrats or Republicans. This stability helps mitigate the risk of underestimating out-party animosity and negative voting.

## 5 Anti-Trump Sentiment

Anti-Trump sentiment arises from three main sources. The first is his anti-immigrant policies, which demonized undocumented immigrants and drew widespread criticism for his handling of race relations. Additionally, the economic recession triggered by the COVID-19 pandemic led to further criticism of Trump's leadership in managing both the economy and the pandemic. The Nationscape data includes four key questions that assess respondents' approval or disapproval of Trump's handling of race relations, the coronavirus outbreak, the economy, and his overall performance as President. Responses were measured on a 4-point Likert scale, ranging from "Strongly Approve" to "Strongly Disapprove" (see the original survey in the Appendix). These questions collectively capture broad negative sentiment toward Trump in 2020, ensuring conceptual alignment.

Given the high correlations among these variables, with a scale reliability coefficient of 0.98 (see the correlation matrix in Table A2 in the Appendix), it is advisable to combine them into a composite measure of "anti-Trump sentiment." This construct reflects overall negative sentiment toward Trump, incorporating views on his handling of the economy, race relations, the pandemic, and his overall presidential leadership.

Figure 1. Anti-Trump Sentiment

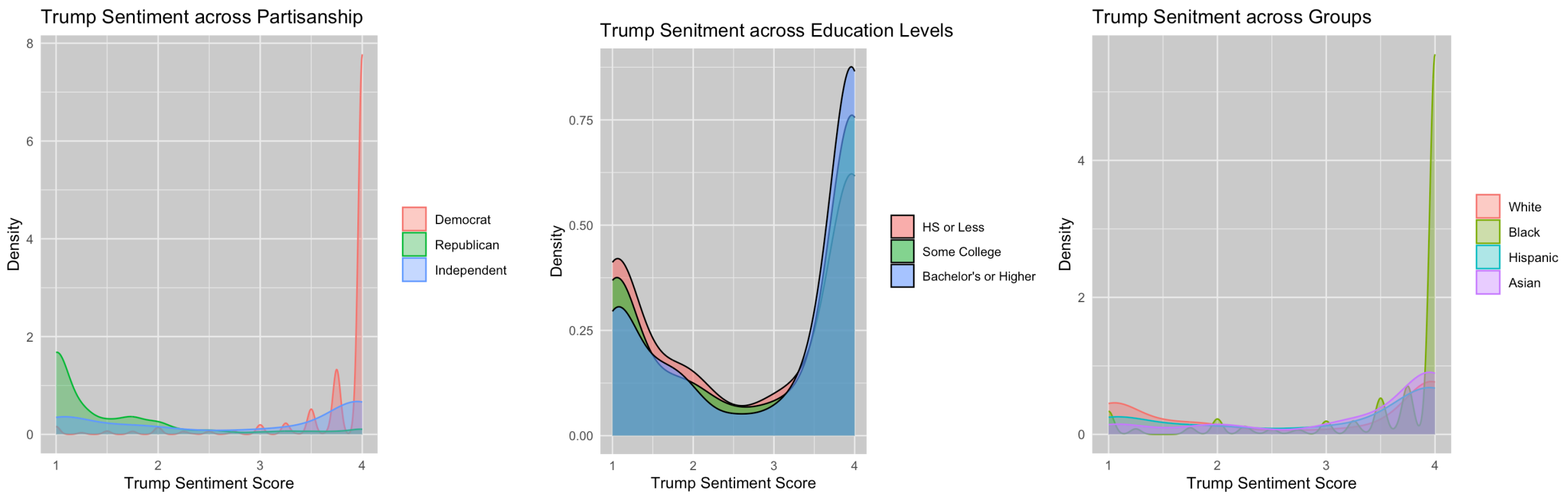

As shown in Figure 1, the anti-Trump sentiment score ranges from 1 to 4, where 1 indicates high approval and 4 indicates high disapproval of Trump's performance. Summary statistics for these variables are provided in Table A1 in the Appendix. Figure 1 illustrates the distribution of anti-Trump sentiment across different demographic groups, including partisanship, education levels, and racial groups. Democrats overwhelmingly disapproved of Trump's leadership, while Republicans generally approved, though less strongly than Democrats opposed him. Independents were more divided but leaned slightly toward disapproval. Individuals with a college degree or higher exhibited significantly higher levels of disapproval. Among racial groups, African Americans expressed the strongest anti-Trump sentiment, with other racial groups also showing more disapproval than approval. Overall, the 2020 election saw a significant surge in anti-Trump sentiment, which played a key role in driving asymmetric negative voting against him.

## 6 Demographic Shifts in Trump Support & Asymmetric Negative Voting

Despite challenges posed by the pandemic—such as obstacles to voter registration and limited access to polling places—the 2020 election saw the highest voter turnout of the 21st century, with a record-breaking increase of 17 million voters compared to 2016 (Fabina, 2021). Trump's declining popularity highlighted significant demographic shifts between the 2016 and 2020 elections. Table 1 presents demographic differences in support for Trump across the two election years.

At first glance, support among Democrats and Republicans showed little change, while independents experienced a notable 7.11 percent decrease in their support for Trump. Among educational groups, those with a high school education or less, as well as individuals with some college education, saw significant decreases in support—11.62 percent and 7.82 percent, respectively. When we examine partisan groups by education level—specifically "high school or less" and "college and higher"—individuals with lower educational attainment exhibited greater variation. For example, among low-education independents, the percentage of votes for Trump increased by 15.26 percent between 2016 and 2020.

Additionally, voters with an income below $40,000 experienced a sharp 16.56 percent decrease in support for Trump, which is notable given that this group had shown higher support levels in 2016. Voters under 40 years old saw an increase in Trump support by 5-8 percent, possibly influenced by the political climate, social movements, and Trump's policies. Despite this shift, these voters may have leaned toward Trump in 2020 due to specific issues like cultural polarization, which might have resonated with them more than with older generations.

Conversely, for voters aged 40 and older, the decline in Trump support can be attributed to factors such as his handling of the pandemic, social justice issues, and his overall leadership style. Older voters, especially those who supported him in 2016, may have become disillusioned, leading to a reduction in their support in 2020.

Table 1. Demographic Shifts in Trump Voter Support

| | 2016 | 2020 | Difference |
|---|---|---|---|
| Democrat | 6.48 | 4.13 | -2.35 |
| Republican | 90.43 | 90.89 | 0.46 |
| Independent | 48.93 | 41.82 | -7.11 |
| Democrat (college or higher) | 3.09 | 2.83 | -0.26 |
| Democrat (HS or less) | 12.64 | 4.78 | -7.86 |
| Republican (college or higher) | 85.68 | 87.72 | 2.04 |
| Republican (HS or less) | 95.05 | 93.27 | -1.78 |
| Independent (college or higher) | 39.95 | 35.65 | -4.3 |
| Independent (HS or less) | 63.47 | 48.21 | -15.26 |
| Men | 48.82 | 44.05 | -4.77 |
| Women | 36.83 | 33.33 | -3.5 |
| White | 50.16 | 46.31 | -3.85 |
| Black | 8.25 | 8.44 | 0.19 |
| Hispanic | 45.7 | 49.06 | 3.36 |
| Asian | 22.58 | 24 | 1.42 |
| HS or Less | 57.87 | 46.25 | -11.62 |
| Some College | 46.63 | 38.81 | -7.82 |
| College or Higher | 38.82 | 33.22 | -5.6 |
| Income (<40K) | 41.18 | 24.62 | -16.56 |
| Income (40-99.99k) | 38.32 | 33.01 | -5.31 |
| Income (>100k) | 45.05 | 38.51 | -6.54 |
| Age <30 | 18.1 | 23.49 | 5.39 |
| Age (31-40) | 25.9 | 32.78 | 6.88 |
| Age (41-50) | 42.81 | 34.6 | -8.21 |
| Age (51-60) | 51.18 | 44.52 | -6.66 |
| Age (61-70) | 49.61 | 43.33 | -6.28 |
| Age (71+) | 45.86 | 42.31 | -3.55 |

*Source:* The 2016 and 2020 Nationscape data

## *6.1 Perception of Economic Change and Negative Partisanship Trend*

Between 2016 and 2020, Americans' perceptions of the economy shifted from optimism to pessimism. This change was closely tied to the onset of the COVID-19 pandemic, which created widespread uncertainty about the future. In the Nationscape dataset, respondents were asked, "Overall, do you think the economy is getting better or worse?" The response options were: "getting better," "about the same," and "getting worse."

As illustrated in Figure 2, most respondents initially believed the economy was either stable or improving before 2020. However, by September 2020, growing uncertainty surrounding the pandemic led to a notable shift in perception, with the majority indicating that they felt the economy was deteriorating.

Figure 2. Perception of Economic Change from 2016 to 2020

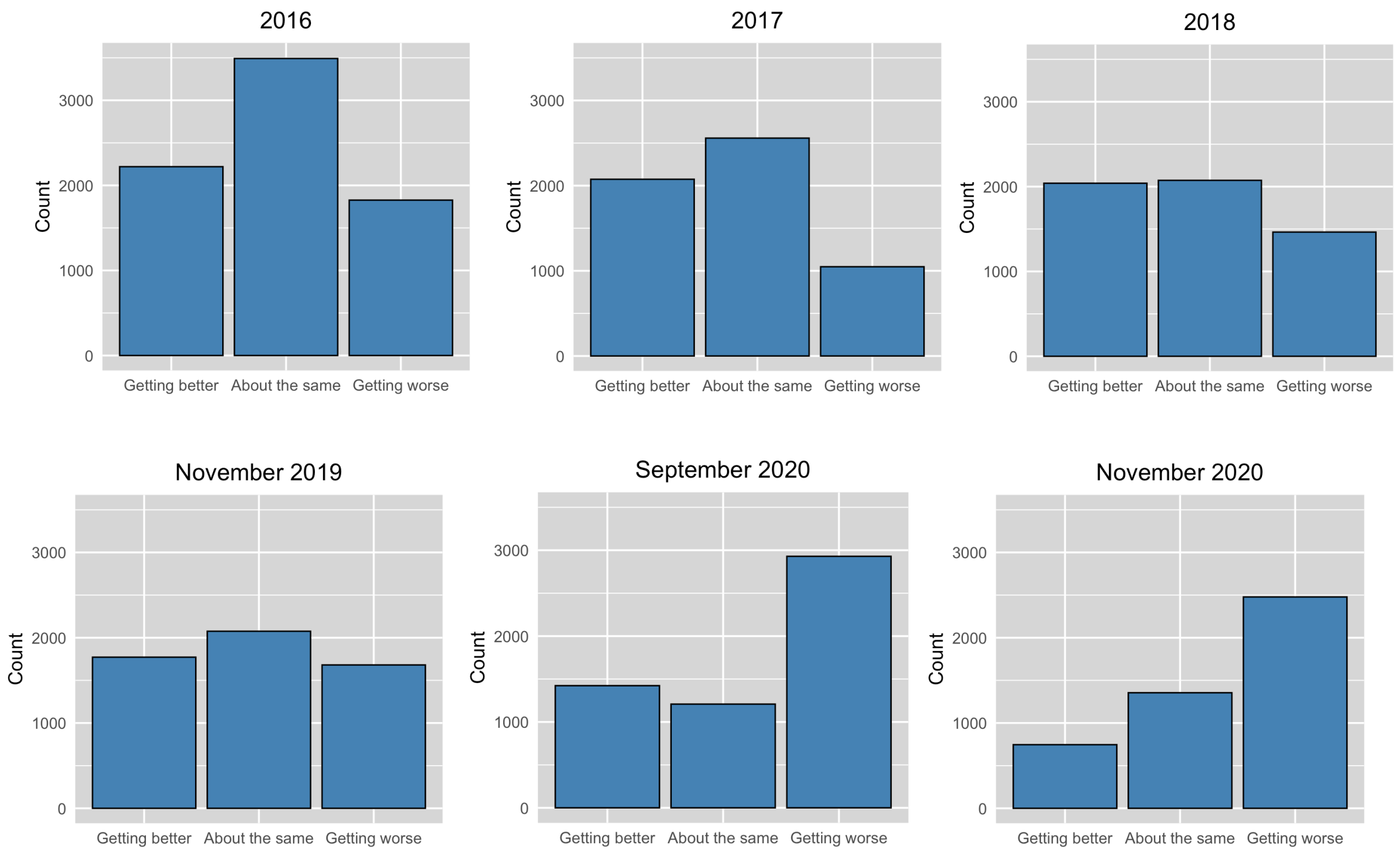

## *6.2 Negative Partisanship*

Alongside economic concerns, negative partisanship surged significantly between 2017 and 2020. The Nationscape data captured respondents' feelings toward Democrats and Republicans, which were used to calculate negative partisanship. Negative partisanship is defined as the difference between feelings toward Democrats and feelings toward Republicans. A positive value indicates a preference for Democrats over Republicans, while a negative value suggests the opposite. If respondents expressed equal feelings toward both parties, their negative partisanship score would be zero. Figure 3 illustrates the trend from 2017 to 2020. In 2017, negative partisanship was relatively balanced between Democrats and Republicans. However, starting in 2019, the trend shifted notably, showing a growing tilt in favor of Democrats.

Figure 3. Negative Partisanship from 2017 to 2020.

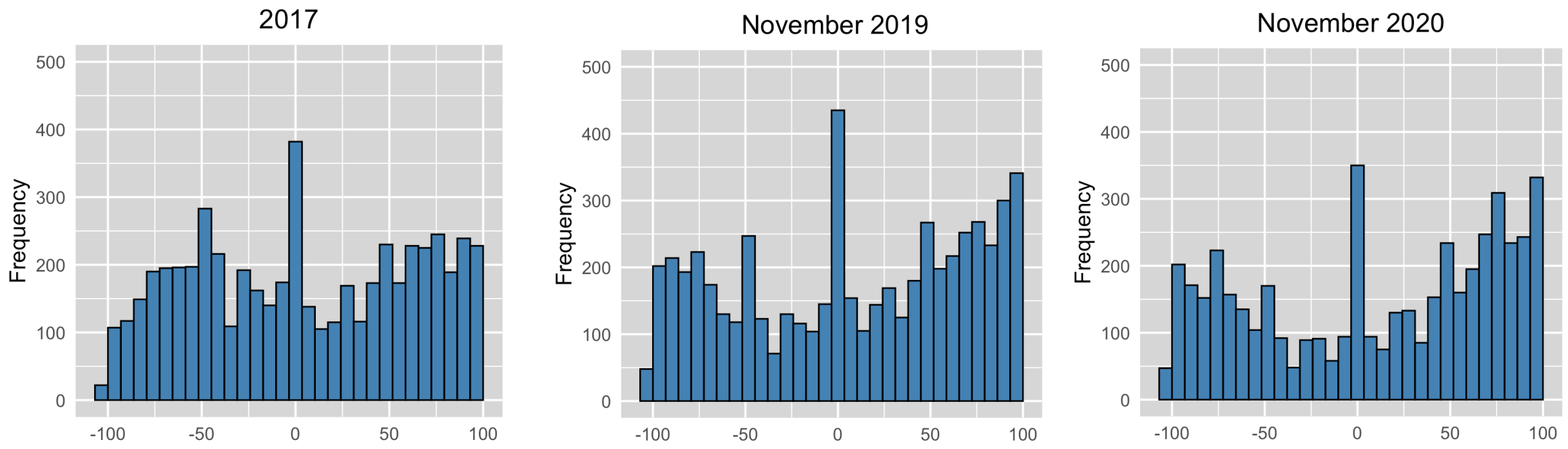

Note: For additional plots, please refer to Figure A2 in the Appendix.

Trump's favorability highlights a stark partisan divide. In 2020, over 80 percent of individuals with a positive view of Trump identified as Republicans, while fewer than 15 percent were Democrats. Among Independents, Trump's favorability remained nearly evenly split from 2017 to 2020, with many adopting more moderate views, neither strongly positive nor strongly negative (See Figure A1 in Appendix). Moreover, partisan strength plays a significant role in negative voting. Echoing the findings of Garzia and Ferreira da Silva (2022), our analysis reveals that individuals with weaker partisan ties were more likely to engage in negative voting in both the 2016 and 2020 elections. Notably, among partisans, Democrats are more inclined to cast negative votes against out-party candidates compared to Republicans (see Figure A3 in the Appendix).

## *6. 3 Negative Voting*

Negative voting was relatively balanced in 2016; however, this pattern shifted significantly in 2020. As shown in Figure 4, approximately 25 percent of voters cast negative votes against Hillary Clinton, while 17 percent did so against Donald Trump in 2016. By the 2020 election, negative voting against Joe Biden dropped to 8 percent, while negative voting against Donald Trump rose significantly to 27 percent. This shift highlights the growing intensity of negative voting directed toward Trump in 2020, particularly among Democratic and Independent voters. In contrast, negative voting against Biden remained relatively low, indicating a less polarized response to the Democratic candidate compared to 2016.

Figure 4. Partisan Animosity and Negative Voting in 2016 and 2020

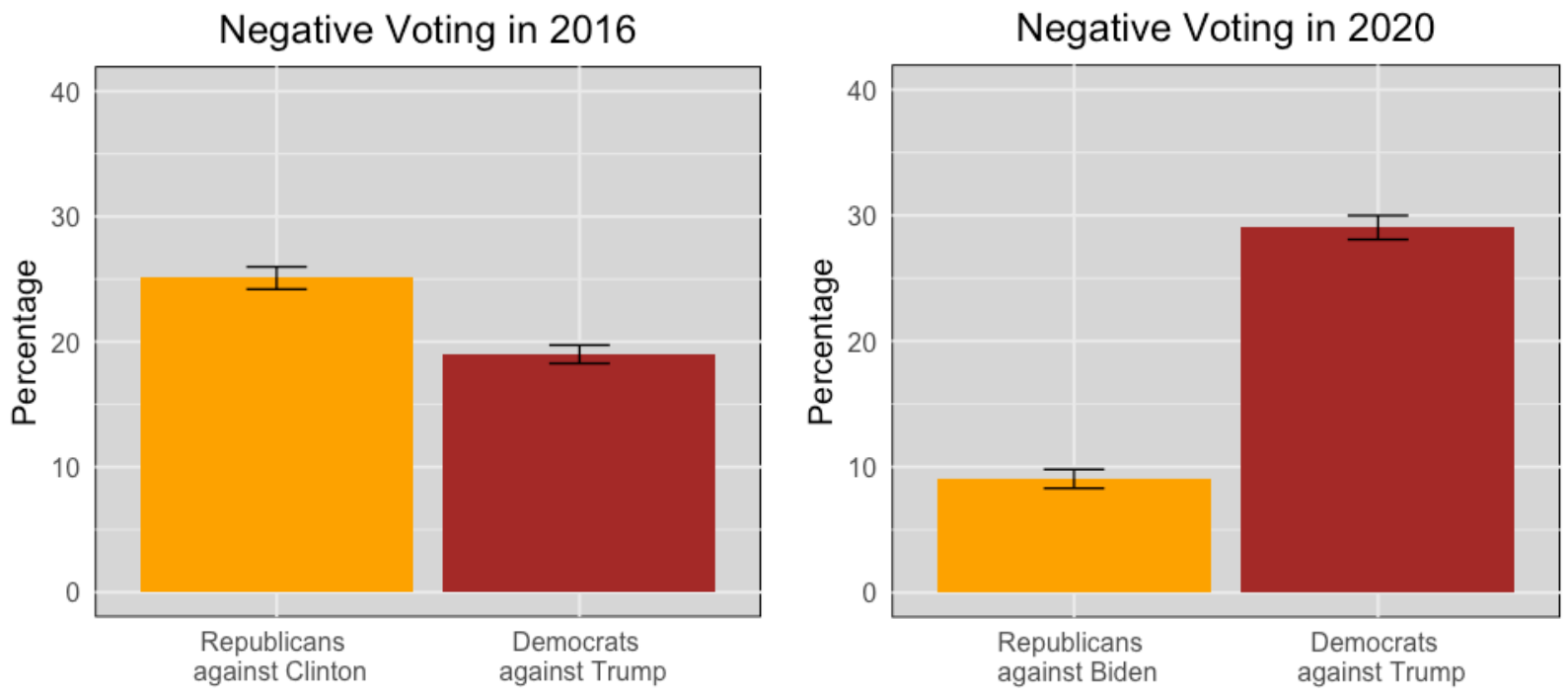

Figure 5. Negative Voting among Independents

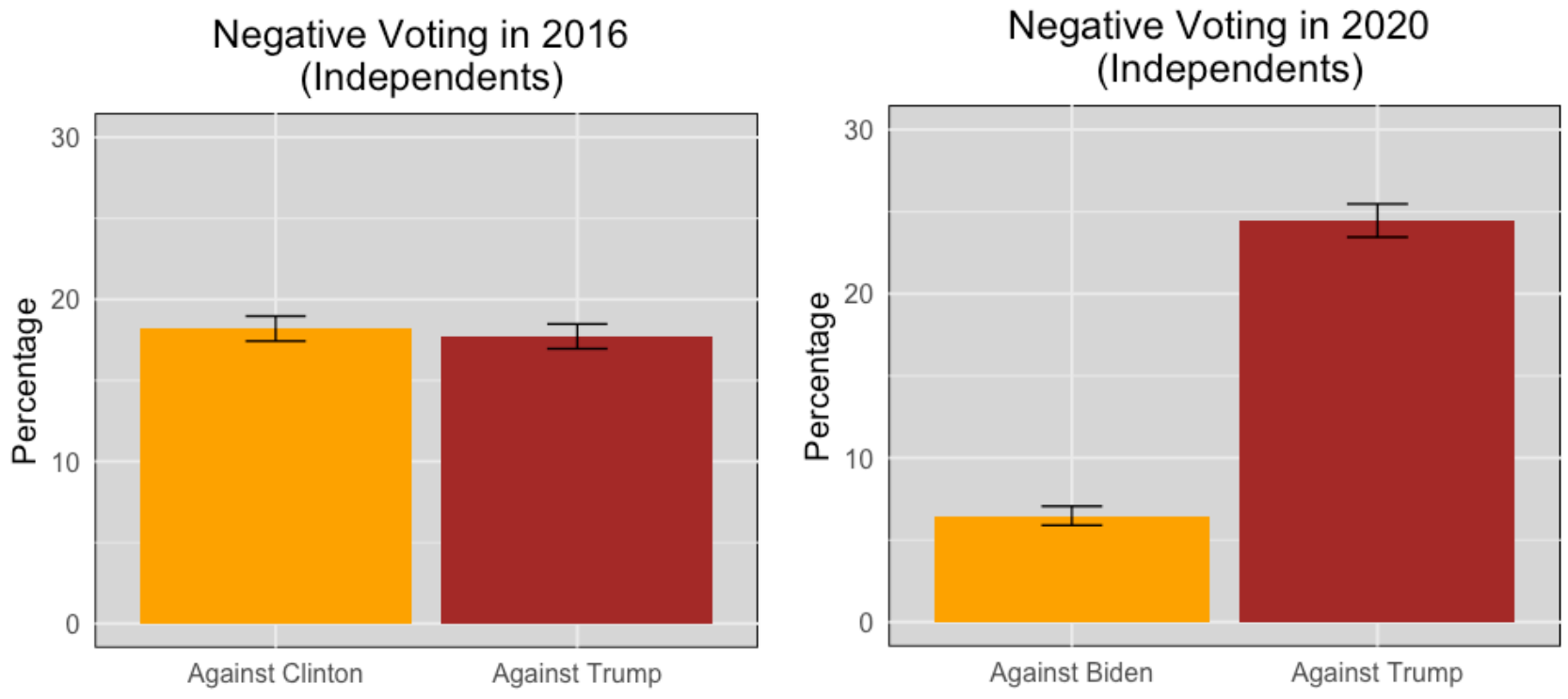

Independents exhibited similar patterns, as shown in Figure 5, with approximately 20 percent engaging in negative voting against both Hillary Clinton and Donald Trump in 2016. However, in 2020, negative voting against Trump among Independents increased to around 25 percent, while negative voting against Joe Biden remained relatively low at about 7 percent. This suggests that the intensity of opposition toward Trump among Independents grew considerably between the two elections. In contrast, favor voting remained relatively symmetric across different partisan groups in both 2016 and 2020 (see Figure A4 in the Appendix), indicating a consistent level of positive support for candidates within each party. This balance in favor voting underscores the distinctiveness of negative partisanship patterns. In summary, the descriptive statistics highlight that negative voting against Trump in the 2020 election was primarily driven by Democrats and Independents, while Republicans largely refrained from negative voting against Biden. These findings illustrate the asymmetric nature of negative partisanship in 2020, with Trump becoming a more polarizing figure compared to Biden.

### *6.4 Negative Voting and Demographic Differences*

Demographic factors such as age and education played a critical role in negative voting during the 2020 election. As shown in Figure 6, in 2016, the age distribution of those who negatively voted against Clinton was relatively normal, with an average age of around 55. In contrast, the age distribution of those who negatively voted against Trump was slightly skewed to the left, as young adults under 35 were more likely to cast negative votes against him. Despite these differences, the distributions in 2016 were relatively comparable. However, the 2020 election marked a stark contrast. The number of respondents negatively voting against Biden was significantly smaller, while negative voting against Trump surged dramatically, particularly among voters under 35.

Figure 6. Age-Based Distributions of Negative Voting in 2016 and 2020

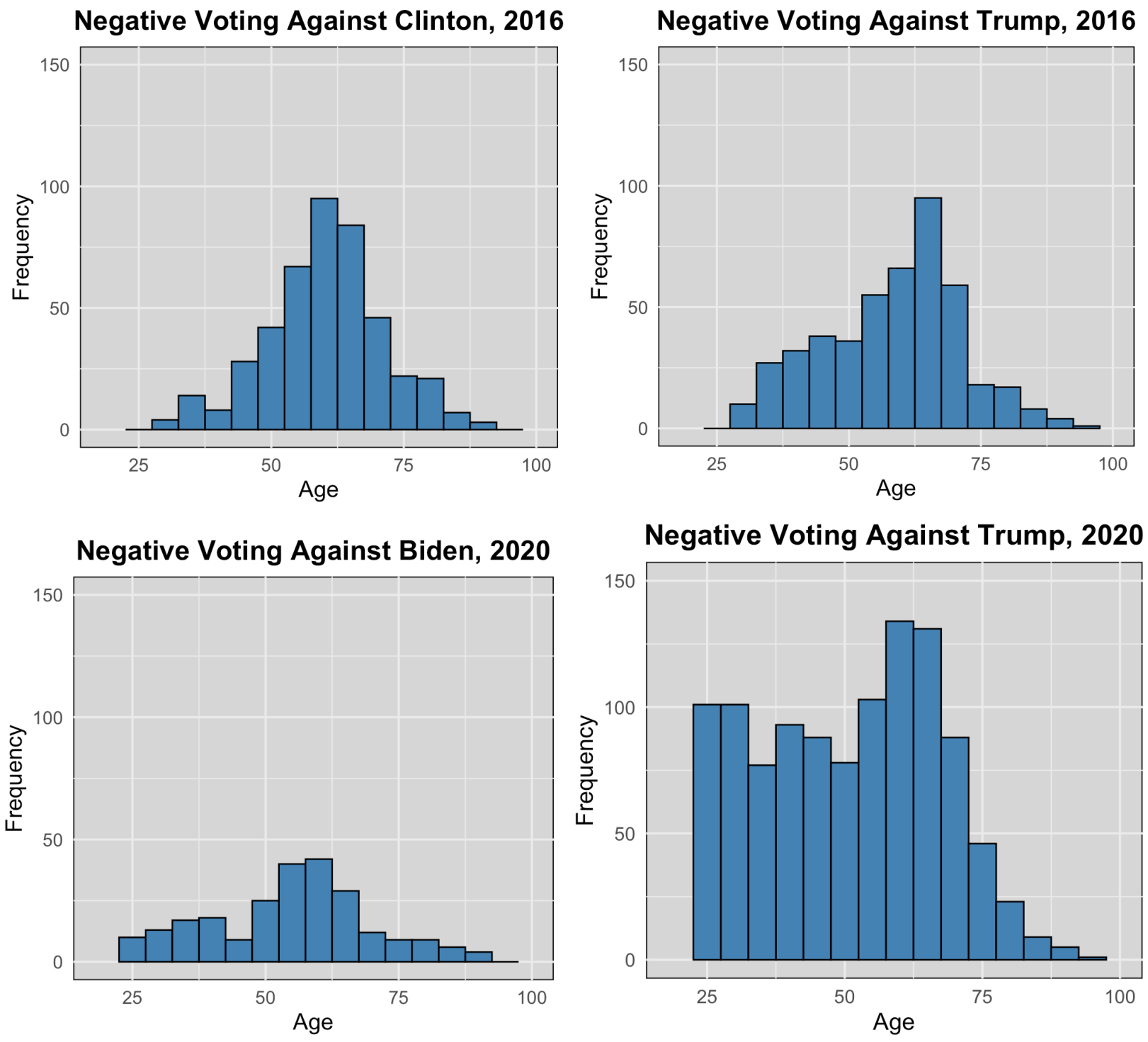

Figure 7. Negative Voting and Education

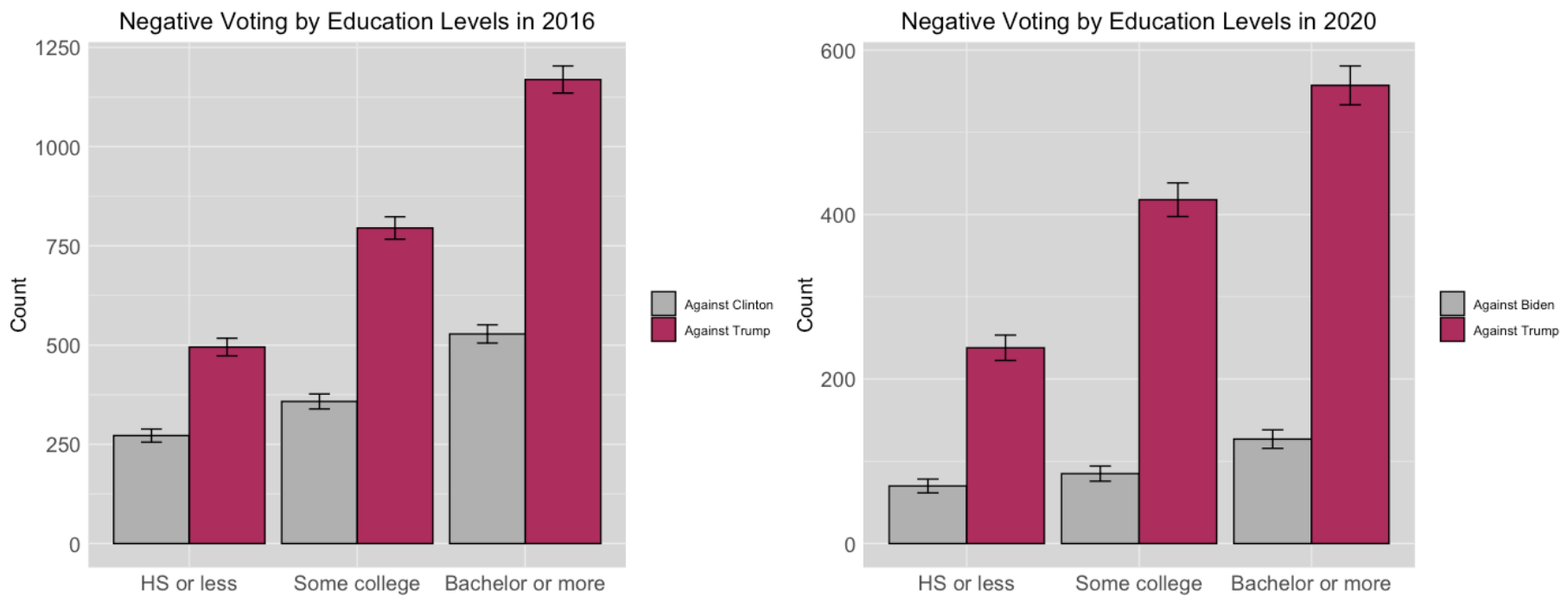

As shown in Figure 7, education played a significant role in negative voting behavior. In 2016, individuals were more than twice as likely to cast a negative vote against Trump compared to Clinton. By 2020, the odds of negative voting against Trump versus Biden had quadrupled, indicating that respondents opposing Trump significantly outnumbered those opposing Biden.

# 7 Statistical Analysis

## *7.1 The 2016 Election*

We start the statistical analysis with the 2016 presidential election. Table 1 presents four logistic regression models where the dependent variables are binary, coded as 1 for either favor voting for or negative voting against Hillary Clinton or Donald Trump, and 0 otherwise. The key independent variables are partisanship, education, race and ethnicity, as well as age. We also include a set of binary control variables, such as indicators for three income brackets, born-again evangelical status, interest in news, gender (female), and homeownership. In Table 2, Models 1 and 2 revealed that Democrats and Republicans exhibited a higher likelihood of voting for their respective in-party candidates, while Independents were more inclined to vote for Hillary Clinton. Moreover, individuals with higher income, advanced educational attainment, and females were more likely to support Clinton based on favorability. On the contrary, motivations for negative voting against the opponent candidates appeared to have a simpler rationale. As demonstrated by Models 3 and 4, in-party affiliation significantly influenced why individuals cast negative votes against out-party candidates. Democrats were more likely to negatively vote against Trump, and similarly, Republicans were inclined to negatively vote against Clinton. However, negative voting among Democrats against Trump in 2016 was barely statistically significant, while among Republican negative voting against Hillary Clinton was statistically significant. This is perhaps because Trump was still relatively new to the American electorate, despite his controversial rhetoric. In contrast, by 2016, Hillary Clinton had been a well-known political figure whom many Republicans disliked. Among independents, they tended to favor Clinton, and negatively voted against Trump in 2016.

Table 2. Determinants of Favor and Negative Voting in 2016

| | (1) | (2) | (3) | (4) |
|---|---|---|---|---|
| **Partisanship** | **Favor Clinton** | **Favor Trump** | **Against Clinton** | **Against Trump** |
| Democrat | 2.848*** | -2.339*** | -1.855*** | 0.393* |
| | (0.222) | (0.173) | (0.222) | (0.201) |
| Republican | -1.840*** | 1.166*** | 0.523*** | -1.654*** |
| | (0.276) | (0.148) | (0.188) | (0.238) |
| Independent | 0.706*** | -0.180 | 0.183 | 0.410** |
| | (0.225) | (0.149) | (0.189) | (0.202) |
| **Education (Reference: HS or less)** | | | | |
| Some college | 0.176** | -0.348*** | 0.0741 | 0.407*** |
| | (0.088) | (0.079) | (0.095) | (0.102) |
| Bachelor or more | 0.473*** | -0.852*** | 0.247*** | 0.443*** |
| | (0.090) | (0.082) | (0.094) | (0.103) |
| **Race/Ethnicity (Reference: Black)** | | | | |
| White | -0.867*** | 0.785*** | 1.579*** | -0.030 |
| | (0.105) | (0.179) | (0.313) | (0.119) |
| Hispanic | -0.462*** | 0.530** | 1.366*** | -0.216 |
| | (0.166) | (0.222) | (0.348) | (0.193) |
| Asian | -1.018*** | 0.434 | 0.756 | 0.417* |
| | (0.251) | (0.338) | (0.487) | (0.244) |
| **Age (Reference: 51-60)** | | | | |
| <30 | 0.458* | -0.906** | -0.815* | 0.346 |
| | (0.275) | (0.359) | (0.442) | (0.280) |
| 31-40 | -0.270 | -0.784*** | -0.499** | 0.651*** |
| | (0.169) | (0.191) | (0.221) | (0.169) |
| 41-50 | 0.082 | -0.407*** | 0.0249 | 0.266* |
| | (0.132) | (0.127) | (0.138) | (0.144) |
| 61-70 | 0.221** | 0.0374 | -0.012 | 0.206* |
| | (0.110) | (0.100) | (0.114) | (0.125) |
| 71+ | 0.110 | -0.165* | -0.090 | 0.257** |
| | (0.092) | (0.084) | (0.096) | (0.105) |
| **Income** | | | | |
| Low (<40k) | 0.072 | -0.005 | -0.379*** | 0.131 |
| | (0.112) | (0.104) | (0.122) | (0.122) |
| Medium (40-99.99k) | 0.048 | 0.017 | -0.023 | 0.046 |
| | (0.103) | (0.093) | (0.103) | (0.111) |
| High (>=100k) | 0.318*** | -0.141 | -0.075 | 0.057 |
| | (0.116) | (0.107) | (0.117) | (0.125) |
| Born Again Evangelical | -0.454*** | 0.312*** | 0.305*** | -0.703*** |
| | (0.084) | (0.067) | (0.076) | (0.100) |
| News Interest | 0.580*** | 0.503*** | 0.104 | 0.063 |
| | (0.108) | (0.110) | (0.131) | (0.120) |
| Married | -0.094 | 0.172** | 0.155* | -0.080 |
| | (0.070) | (0.069) | (0.080) | (0.076) |
| Female | 0.227*** | -0.131 | -0.191* | 0.077 |
| | (0.084) | (0.086) | (0.100) | (0.092) |
| Home Ownership | 0.019 | -0.018 | 0.036 | 0.002 |
| | (0.078) | (0.074) | (0.084) | (0.085) |
| Intercept | -2.316*** | -1.516*** | -3.421*** | -2.403*** |
| | (0.291) | (0.268) | (0.397) | (0.290) |
| N | 7,599 | 7,599 | 7,599 | 7,599 |

Standard errors in parentheses, *** p<0.01, ** p<0.05, * p<0.1

## *7.2 The 2020 Election*

Table 3 presented four logistic regression models where the dependent variables were binary, coded as 1 for either favor voting for or negative voting against Joe Biden or Donald Trump, and 0 otherwise.

Table 3 focuses on broad socioeconomic factors influencing the likelihood of favoring or negatively voting against Biden and Trump. Compared to the 2016 election, the 2020 election was more dynamic. In general, partisanship remained the most important predictor for voting against opponent candidates. For Democrats, voting against Donald Trump was a major motivation for many to vote for Joe Biden. Similarly, voting against Joe Biden motivated many Republicans to vote for Donald Trump. Unlike in the 2016 election, voting against Trump was statistically significant among Democrats in 2020, indicating that animosity toward Donald Trump had increased substantially among them. Most importantly, aversion to Trump became a significant motivation for Independents to negatively vote against Trump and support Biden.

The regression results in Table 3 showed that Democrats and Republicans generally favored their own party's candidates and opposed the opposing party's candidates. However, while some Independents supported Trump in 2020, there was a clear trend of Independents voting negatively for Biden, primarily due to their opposition to Trump. Moreover, individuals who negatively voted against Trump were more likely to be college-educated and White. Younger voters, particularly those under 30, were also more likely to have voted negatively against Trump. People with higher incomes and those with greater interest in news were more inclined to vote negatively against Trump.

Table 3. Vote in Favor or Against Candidates in 2020.

| | (1) | (2) | (3) | (4) |
|---|---|---|---|---|
| **Partisanship** | **Favor Biden** | **Favor Trump** | **Against Biden** | **Against Trump** |
| Democrat | 2.591*** | -2.358*** | -1.745*** | 0.567*** |
| | (0.267) | (0.249) | (0.393) | (0.168) |
| Republican | -1.357*** | 2.288*** | 0.710** | -1.765*** |
| | (0.328) | (0.205) | (0.325) | (0.217) |
| Independent | 0.823*** | 0.378* | 0.391 | 0.350** |
| | (0.274) | (0.203) | (0.322) | (0.171) |
| **Education (Reference: HS or less)** | | | | |
| Some college | 0.067 | 0.124 | 0.0675 | 0.240** |
| | (0.102) | (0.113) | (0.179) | (0.100) |
| Bachelor or more | 0.077 | -0.361*** | 0.375** | 0.455*** |
| | (0.108) | (0.120) | (0.179) | (0.104) |
| **Race/Ethnicity (Reference: Black)** | | | | |
| White | -0.619*** | 1.257*** | 0.637* | 0.085 |
| | (0.119) | (0.227) | (0.356) | (0.119) |
| Hispanic | -0.381** | 1.238*** | -0.013 | -0.286* |
| | (0.153) | (0.259) | (0.431) | (0.156) |
| Asian | -0.694*** | 0.670* | 0.285 | 0.128 |
| | (0.246) | (0.360) | (0.529) | (0.219) |
| **Age (Reference: 51-60)** | | | | |
| <30 | -0.736*** | -1.080*** | -0.057 | 0.682*** |
| | (0.142) | (0.166) | (0.245) | (0.125) |
| 31-40 | -0.326** | -0.661*** | 0.248 | 0.328*** |
| | (0.137) | (0.158) | (0.216) | (0.126) |
| 41-50 | 0.0113 | -0.369** | -0.0799 | 0.196 |
| | (0.132) | (0.149) | (0.222) | (0.125) |
| 61-70 | 0.287** | 0.070 | -0.271 | -0.133 |
| | (0.112) | (0.121) | (0.190) | (0.111) |
| 71+ | 0.317** | -0.051 | 0.143 | -0.065 |
| | (0.136) | (0.152) | (0.220) | (0.136) |
| **Income** | | | | |
| Low (<40k) | 0.037 | -0.233 | -0.088 | -0.018 |
| | (0.136) | (0.153) | (0.244) | (0.130) |
| Medium (40-99.99k) | -0.075 | -0.207 | 0.059 | 0.191 |
| | (0.129) | (0.140) | (0.216) | (0.121) |
| High (>=100k) | -0.177 | -0.373** | 0.309 | 0.380*** |
| | (0.146) | (0.160) | (0.234) | (0.136) |
| Born Again Evangelical | -0.155 | 0.757*** | 0.0643 | -0.637*** |
| | (0.102) | (0.099) | (0.147) | (0.103) |
| News Interest | 0.379*** | 0.403*** | -0.222 | 0.400*** |
| | (0.117) | (0.129) | (0.188) | (0.111) |
| Married | -0.027 | 0.165* | 0.170 | -0.100 |
| | (0.088) | (0.098) | (0.150) | (0.084) |
| Female | 0.267*** | -0.255*** | -0.205 | -0.007 |
| | (0.079) | (0.089) | (0.134) | (0.074) |
| Home Ownership | 0.075 | 0.282** | -0.030 | -0.052 |
| | (0.092) | (0.111) | (0.170) | (0.086) |
| Intercept | -2.280*** | -2.599*** | -3.432*** | -1.963*** |
| | (0.333) | (0.347) | (0.538) | (0.257) |
| N | 4,713 | 4,713 | 4,713 | 4,713 |

Standard errors in parentheses, *** p<0.01, ** p<0.05, * p<0.1

## *7.3 The Effects of Anti-Trump Sentiment*

To better understand negative voting, Table 4 incorporates anti-Trump sentiment and presents separate analyses by partisan groups, assessing the effects of anti-Trump sentiment, negative partisanship, and perceived economic decline. To address multicollinearity concerns between partisanship and anti-Trump sentiment, the analysis was conducted on separate data subsets for Democrats, Republicans, and Independents. Given the high correlation ($r = 0.9$) between negative partisanship and anti-Trump sentiment, these variables were analyzed in separate models. Models 1, 3, and 5 focus on anti-Trump sentiment, while Models 2, 4, and 6 examine negative partisanship. Each subset includes the same covariates as in previous models for consistency.

The results indicate that anti-Trump sentiment significantly influenced negative voting against Trump. Among Democrats, the effect was smallest, as their baseline opposition to Trump was already high. Since many Democrats were already likely to vote against him, disapproval of his policies did not substantially increase their likelihood of doing so, leading to a smaller marginal effect. In contrast, Republicans who disapproved of Trump were far more likely to break with partisan loyalty and vote against him, resulting in a larger coefficient. This shift reflects the weight of anti-Trump sentiment among Republicans, as it represented a major departure from expected partisan behavior. For Independents, the effect of anti-Trump sentiment was moderate, falling between that of Democrats and Republicans. Independents, being less polarized, often held mixed or moderate political views, making their disapproval of Trump's handling of key issues less predictive of their vote choice. While anti-Trump sentiment influenced their decisions, its impact was not as pronounced as it was for Republicans.

Table 4. Effects of Anti-Trump Sentiment on Negative Voting Against Trump

| | (1) | (2) | (3) | (4) | (5) | (6) |
|---|---|---|---|---|---|---|
| | Democrat | | Republican | | Independent | |
| Anti Trump Sentiment | 0.789*** | | 2.732*** | | 1.791*** | |
| | (0.153) | | (0.314) | | (0.144) | |
| Negative Partisanship | | 0.000 | | 0.040*** | | 0.025*** |
| | | (0.002) | | (0.005) | | (0.002) |
| Perceived Economic Trend | 0.175* | 0.316*** | 0.057 | 0.259 | 0.271** | 0.543*** |
| | (0.097) | (0.095) | (0.343) | (0.275) | (0.136) | (0.124) |
| **Education (Reference: HS or less)** | | | | | | |
| Some college | 0.008 | -0.019 | 1.980*** | 1.538*** | 0.198 | 0.105 |
| | (0.145) | (0.144) | (0.685) | (0.596) | (0.213) | (0.207) |
| Bachelor or more | 0.076 | 0.0458 | 1.698*** | 1.561*** | 0.410* | 0.478** |
| | (0.149) | (0.149) | (0.650) | (0.580) | (0.221) | (0.215) |
| **Race/Ethnicity (Reference: Black)** | | | | | | |
| White | 0.376** | 0.299* | 2.534** | 2.544* | 0.170 | -0.020 |
| | (0.156) | (0.155) | (1.267) | (1.447) | (0.266) | (0.272) |
| Hispanic | -0.002 | -0.109 | 3.529** | 3.065* | -0.765** | -0.885** |
| | (0.205) | (0.204) | (1.461) | (1.584) | (0.359) | (0.367) |
| Asian | 0.175 | 0.083 | 3.105** | 3.421** | 0.303 | 0.245 |
| | (0.317) | (0.318) | (1.398) | (1.567) | (0.474) | (0.469) |
| **Age (Reference: 51-60)** | | | | | | |
| <30 | 0.832*** | 0.814*** | -0.967 | -0.651 | 0.396 | 0.731*** |
| | (0.182) | (0.183) | (0.966) | (0.812) | (0.269) | (0.261) |
| 31-40 | 0.253 | 0.205 | -1.328* | -0.247 | 0.494* | 0.654** |
| | (0.188) | (0.187) | (0.793) | (0.660) | (0.279) | (0.266) |
| 41-50 | -0.055 | -0.027 | 0.194 | 0.147 | 0.513* | 0.528** |
| | (0.179) | (0.180) | (0.760) | (0.630) | (0.266) | (0.249) |
| 61-70 | -0.411*** | -0.395** | 0.463 | 0.273 | 0.078 | 0.014 |
| | (0.158) | (0.158) | (0.686) | (0.551) | (0.228) | (0.220) |
| 71+ | -0.197 | -0.216 | -0.395 | 0.189 | -0.149 | -0.205 |
| | (0.184) | (0.184) | (0.782) | (0.705) | (0.296) | (0.292) |
| N | 1,778 | 1,757 | 1,073 | 1,069 | 1,301 | 1,279 |

Standard errors in parentheses, *** p<0.01, ** p<0.05, * p<0.1

Note: See Table A3 for complete results.

Furthermore, Figure 8 shows the marginal effects of anti-Trump sentiment on negative voting against Trump. As illustrated, when the anti-Trump sentiment score reaches its maximum, the probability of negative voting against Trump is approximately 0.6 across Democrats, Republicans, and Independents.

Figure 8. Effect of Anti-Trump Sentiment on Negative Voting

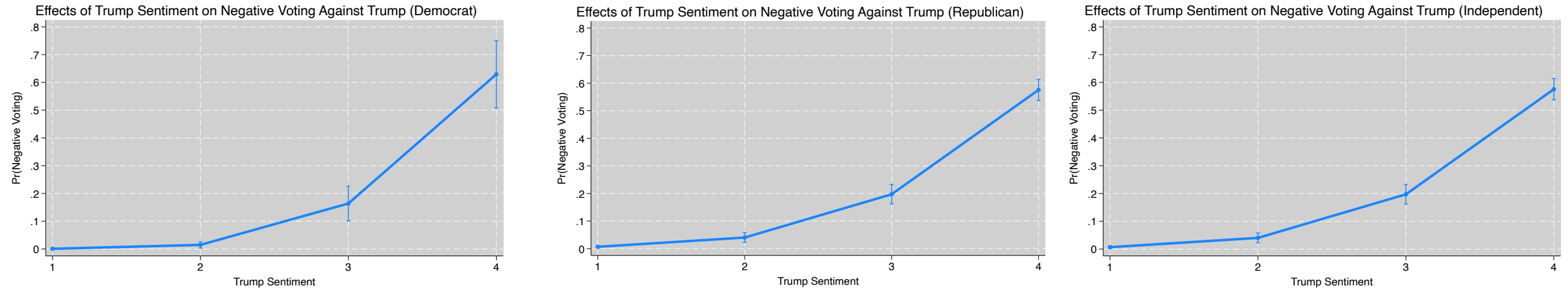

The effects of negative partisanship were similar to those of anti-Trump sentiment, with one key exception: negative partisanship among Democrats was overwhelmingly high, resulting in minimal variation and a lack of statistical significance. Moreover, as shown in Model 4, the statistically significant coefficient of 0.04 indicates that Republicans with positive views toward Democrats were more likely to cast a negative vote against Trump. Independents followed a similar trend, as reflected by the statistically significant coefficient of 0.025, though to a lesser extent.

Nonetheless, perceptions of the economic trend varied slightly across partisan groups. As Models 1 and 2 in Table 4 show, Democrats who perceived a negative economic trend were more likely to cast a negative vote against Trump, as indicated by the statistically significant coefficients of 0.175 and 0.316. However, this variable had no statistically significant effect among Republicans, as shown in Models 3 and 4. Instead, the perception of a negative economic trend had the strongest effect on independents, as demonstrated by the statistically significant coefficients of 0.271 and 0.543 in Models 5 and 6, suggesting that economic concerns were the most significant driver of their negative votes against Trump.

Additionally, younger Democrats, particularly those aged 30 or below, were especially prone to negative voting against Trump. Among Republicans, educated Whites, Latinos, and Asians were the most likely to engage in negative voting. College-educated Independents were also more inclined to cast a negative vote against Trump, whereas Latino Independents were less likely to do so. These nuanced findings highlight the interplay between anti-Trump sentiment, partisanship, and sociodemographic factors in shaping electoral behavior (see Table A3 in the Appendix for full results).

As illustrated in Figure 9, age significantly influences negative voting against Donald Trump among Democrats. Individuals aged 30 or younger were about 50 percent more likely to cast a negative vote against Trump in 2020, while those aged 31 to 40 were 40 percent more likely. However, this effect is minimal among Republicans and shows a not statistically significant correlation among Independents in Table 4.

Figure 9. Effect of Age on Negative Voting

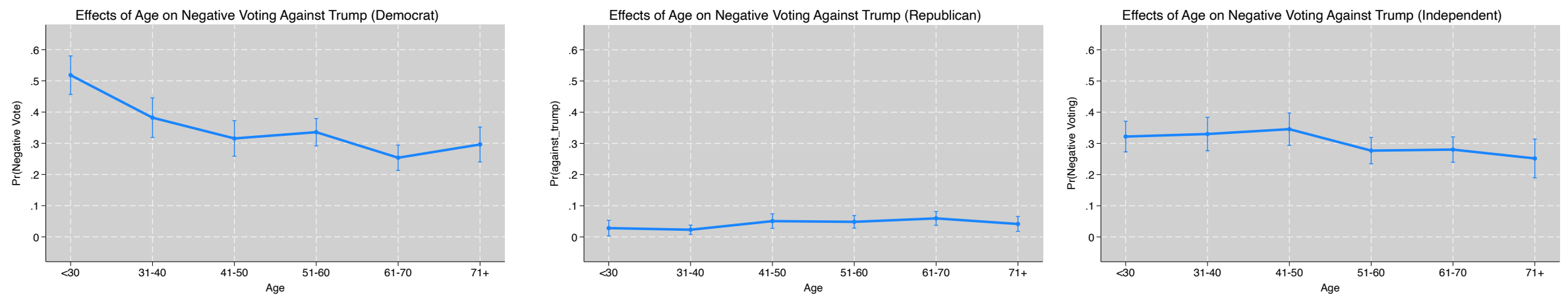

## 8 Discussion & Conclusion

Political scientists have long documented rising polarization in the United States (Hetherington, 2001; Zaller, 1992), yet this study demonstrates how contextual factors—such as elite polarization and crisis management—can drive asymmetry in negative voting, with Trump's presidency serving as a particularly stark example. While previous elections exhibited relatively balanced levels of negative voting, the 2020 election marked a sharp departure, with anti-Trump sentiment driving unprecedented levels of negative voting among young, educated Democrats and Independents. This asymmetry challenges the prevailing assumption that negative partisanship operates uniformly across elections and instead underscores the role of candidate-specific and contextual factors in shaping electoral behavior.

As an incumbent seeking re-election amid economic turmoil, a global pandemic, and widespread political instability, Trump became the focal point of voter dissatisfaction. His handling of race relations, the COVID-19 crisis, and his leadership style reinforced negative perceptions, fueling strong out-party animosity and motivating turnout against him. While Trump's presidency intensified negative partisanship, broader structural shifts—such as changing media consumption patterns and ideological sorting—may have also contributed to the observed asymmetry. In contrast, Republicans demonstrated significantly lower levels of hostility toward Biden than they had toward Clinton in 2016, underscoring the distinct partisan asymmetry of 2020. From a retrospective voting perspective, voters often punish incumbents for perceived failures, but under conditions of extreme elite polarization, these retrospective evaluations become even more potent, amplifying asymmetry in negative voting. While previous studies have treated negative partisanship as a stable, symmetric feature of partisan competition, the 2020 election demonstrates that it is highly contingent on political context, media narratives, and the salience of an incumbent's performance.

These findings contribute to research on affective polarization and retrospective voting in three key ways. First, they provide empirical evidence that negative voting is not always symmetric but can become lopsided under extreme polarization. Second, they show how retrospective evaluations—particularly dissatisfaction with Trump—disproportionately mobilized one side of the electorate, rather than producing balanced negative voting across party lines. Finally, they highlight the role of demographic divides in shaping asymmetric negative voting, with younger and more educated voters far more likely to cast ballots against Trump than Republicans were against Biden. These findings suggest that asymmetric negative voting arises not just from candidate characteristics but also from elite signaling, partisan media reinforcement, and demographic shifts that reshape political identities over time.

Looking ahead, future elections will test whether this pattern of asymmetric negative voting persists beyond Trump's presidency or was a product of his uniquely polarizing leadership. If similar levels of elite polarization and strong retrospective disapproval arise—whether due to economic conditions, governance failures, or deep-seated partisan animosities—negative voting could remain asymmetrical in future contests. If this trend continues, parties may increasingly rely on out-group animosity rather than policy appeals to mobilize voters, further entrenching affective polarization and limiting cross-partisan persuasion. However, it remains possible that 2020 was an anomalous election rather than evidence of a sustained shift in negative voting patterns. Ultimately, future elections will determine whether asymmetric negative voting has become a structural feature of American politics or was uniquely shaped by Trump's presidency.

## References

Abutaleb, Yasmeen, Dawsey Josh, Nakashim Ellen, & Miller Greg. (2020, April 11). The U.S. Was Beset by Denial and Dysfunction as the Coronavirus Raged. *Washington Post*. https://www.washingtonpost.com/national-security/2020/04/04/coronavirus-government-dysfunction/

Beer, Tommy. (2020, December). November's Grim Covid-19 Totals: More Than 4.3 Million Infections And 37,000 Americans Killed. *Forbes*. https://www.forbes.com/sites/tommybeer/2020/12/01/novembers-grim-covid-19-totals-more-than-43-million-infections-and-37000-americans-killed/

Bekafigo, Marija, Stepanova Elena V., Eiler Brian A., Noguchi Kenji, & Ramsey Kathleen. (2019). The Effect of Group Polarization on Opposition to Donald Trump. *Political Psychology*, *40*(5), 1163-1178.

Bennett, Brian, & Rogers B. Tessa. (2020, November 7). How Donald Trump Lost The Election. *Time*. https://time.com/5907973/donald-trump-loses-2020-election/

Bloom, Howard S., & Price H. Douglas. (1975). Voter Response to Short-run Economic Condition. *American Journal of Political Science*, *69*, 1240-1254.

Campbell, Angus, Converse Philip, Miller Warren, & Stokes Donald. (1960). *The American Voter*. John Wiley & Sons, Inc.

Canipe, Chris, Bhatia Gurman, Lange Jason, & Heath Brad. (2020, November 7). How Joe Biden won the U.S. presidential election:. *Reuters Graphic*. https://www.reuters.com/graphics/USA-ELECTION/RESULTS/jznvnjyjzvl/

Democracy Fund Voter Study Group. (2021). *Views of the Electorate Research Survey* (https://www.voterstudygroup.org/

Diamond, Jeremy. (2017, January 29). Trump's latest executive order: Banning people from 7 countries and more. *CNN*. https://www.cnn.com/2017/01/27/politics/donald-trump-refugees-executive-order/index.html

Druckman, James, Levendusky Matthew, & Mclain A. (2018). No Need to Watch: How the Effects of Partisan Media can Spread via Interpersonal Discussion. *American Journal of Political Science*, *61*(1), 99-112.

Fabina, Jacob. (2021). *Despite Pandemic Challenges, 2020 Election Had Largest Increase in Voting Between Presidential Elections on Record*. U. S. C. Bureau. https://www.census.gov/library/stories/2021/04/record-high-turnout-in-2020-general-election.html

Fiorina, Morris P., & Shepsle Kenneth A. (1989). Is Negative Voting an Artifact? *American Journal of Political Science*, *33*(2), 423-439. https://doi.org/https://doi.org/10.2307/2111154

Gant, Michael M., & Davis Dwight F. (1984). Negative Voter Support in Presidential Elections. *The Western Political Quarterly*, *37*(2). https://doi.org/https://doi.org/10.2307/448569

Garzia, Diego, & Ferreira Da Silva Frederico. (2022). The Electoral Consequences of Affective Polarization? Negative Voting in the 2020 US Presidential Election. *American Politics Research*, *50*(3), 303-311. https://doi.org/10.1177/1532673X221074633

Hart, Roderick P. (2022). Why Trump Lost and How? A Rhetorical Explanation. *American Behavioral Scientist*, *66*(1), 7-27. https://doi.org/10.1177/0002764221996760

Hetherington, Marc. (2001). Resurgent Mass Partisanship: The Role of Elite Polarization. *The American Political Science Review*, *95*(No. 3).

Igielnik, Ruth, Keeter Scott, & Hartig Hannah. (2021, June 30). Behind Biden's 2020 Victory: An Examination of the 2020 Electorate, Based on Validated Voters. *Pew Research Center*. https://www.pewresearch.org/politics/2021/06/30/behind-bidens-2020-victory/

Iyengar, Shanto, Sood G., & Lelkes Y. (2012). Affect, Not Ideology: A Social Identity Perspective on Polarization. *Public Opinion Quarterly*, *76*(3), 405-431.

Jacobson, Gary C. (2020). Divide et Impera: Polarization in Trump's America. In M. D. Pero & P. Magr (Eds.), *Four Years of Trump: The U.S. and the World* (pp. 17– 49). Instituto per gli Studi di Politica Internazional.

Jacobson, Gary C. . (2021). Driven to Extremes: Donald Trump's Extraordinary Impact on the 2020 Election. *Presidential Studies Quarterly*, *51*(3), 492-521. https://doi.org/10.1111/psq.12724

Kernell, Samuel. (1977). Presidential Popularity and Negative Voting. *American Political Science Review*, *71*, 44-66.

Macwilliams, Matthew C. (2016). Who Decides When the Party Doesn't? Authoritarian Voters and the Rise of Donald Trump. *PS: Political Science and Politics*, *49*(4).

Mccarty, Nolan, Poole Keith, & Rosenthal Howard. (2016). *Polarized America: The Dance of Ideological and Unequal Riches*. MIT Press.

Pew Research Center. (2020a). *In Changing U.S. Electorate, Race and Education Remain Stark Dividing Lines.*

Pew Research Center. (2020b). *Most Americans Say Trump Was Too Slow in Initial Response to Coronavirus Threat*. https://www.pewresearch.org/politics/2020/04/16/trumps-handling-of-coronavirus-outbreak/

Robison, Joshua, & Moskowitz Rachel L. (2019). The Group Basis of Partisan Affective Polarization. *Journal of Politics*, *81*(3), 1075-1079. https://doi.org/http://dx.doi.org/10.1086/703069

Suhay, E. (2015). Explaining Group Influence: The Role of Identity and Emotion in Political Conformity and Polarization. *Political Behavior*, *37*(1), 221-251.

Tyler, Matthew, & Iyengar Shanto. (2023). Testing the Robustness of the ANES Feeling Thermometer Indicators of Affective Polarization. *American Political Science Review*, *First View*, 1-7. https://doi.org/https://doi.org/10.1017/S0003055423001302

Weber, Christopher, & Klar Samara. (2019). Exploring the Psychological Foundations of Ideological and Social Sorting. *Political Psychology*, *40*. https://doi.org/https://doi.org/10.1111/pops.12574

Zaller, John. (1992). *The Nature and Origins of Mass Opinion*. Cambridge University Press.

# Appendix

Figure A1. Favorability of Trump from 2017 to 2020

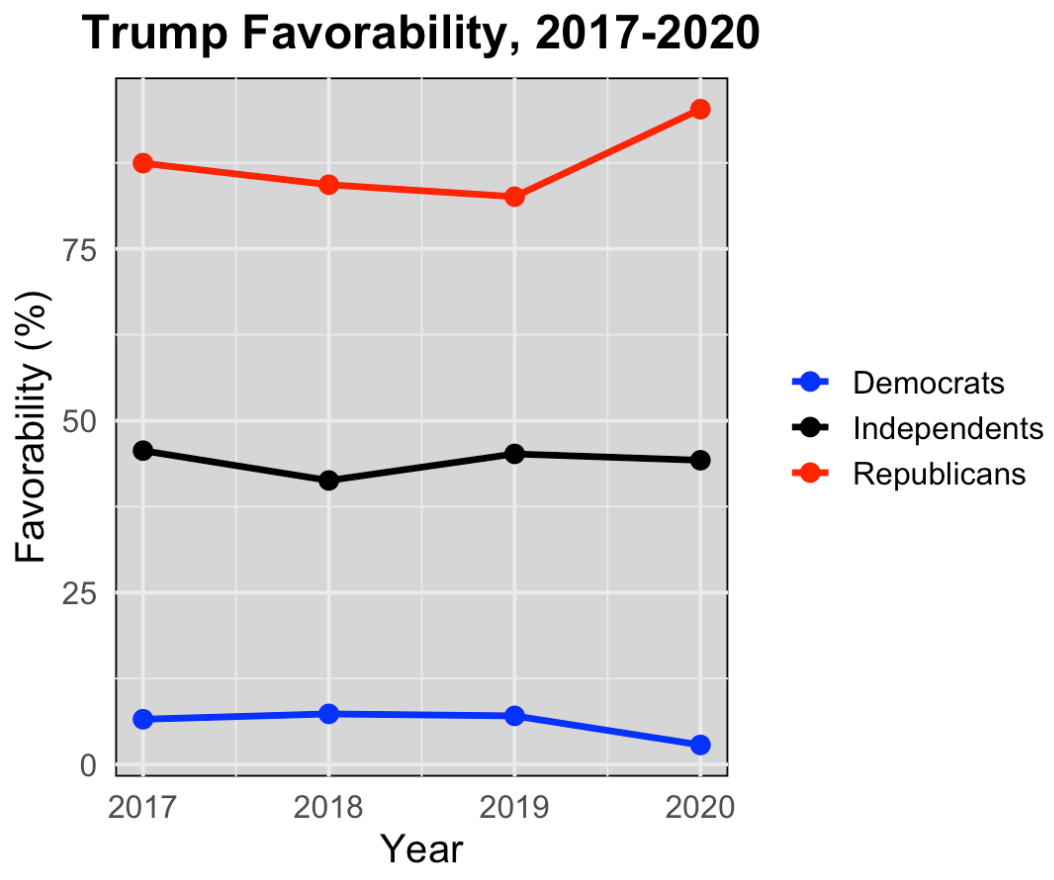

Figure A2. Negative Partisanship in 2019 to 2020.

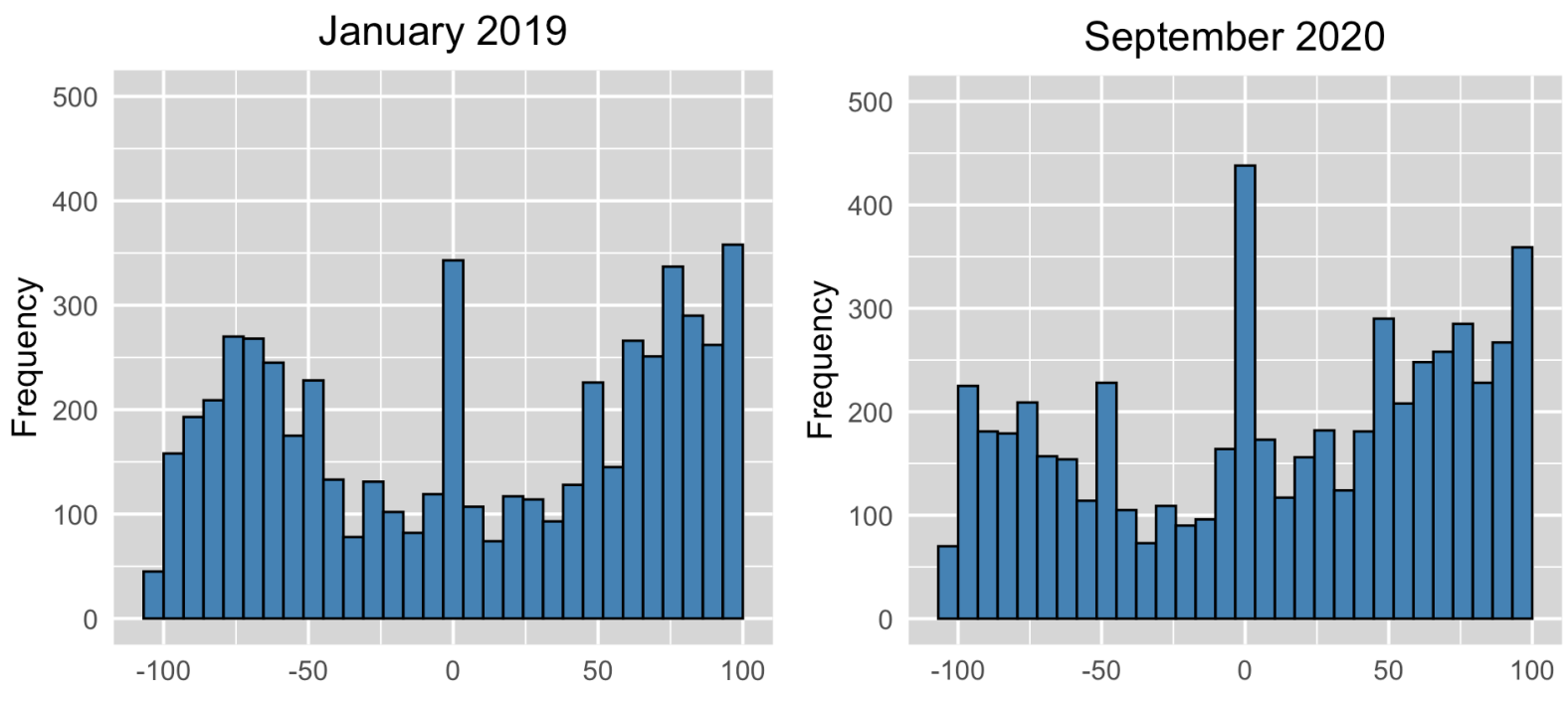

Figure A3. Negative Voting and Partisan Strength

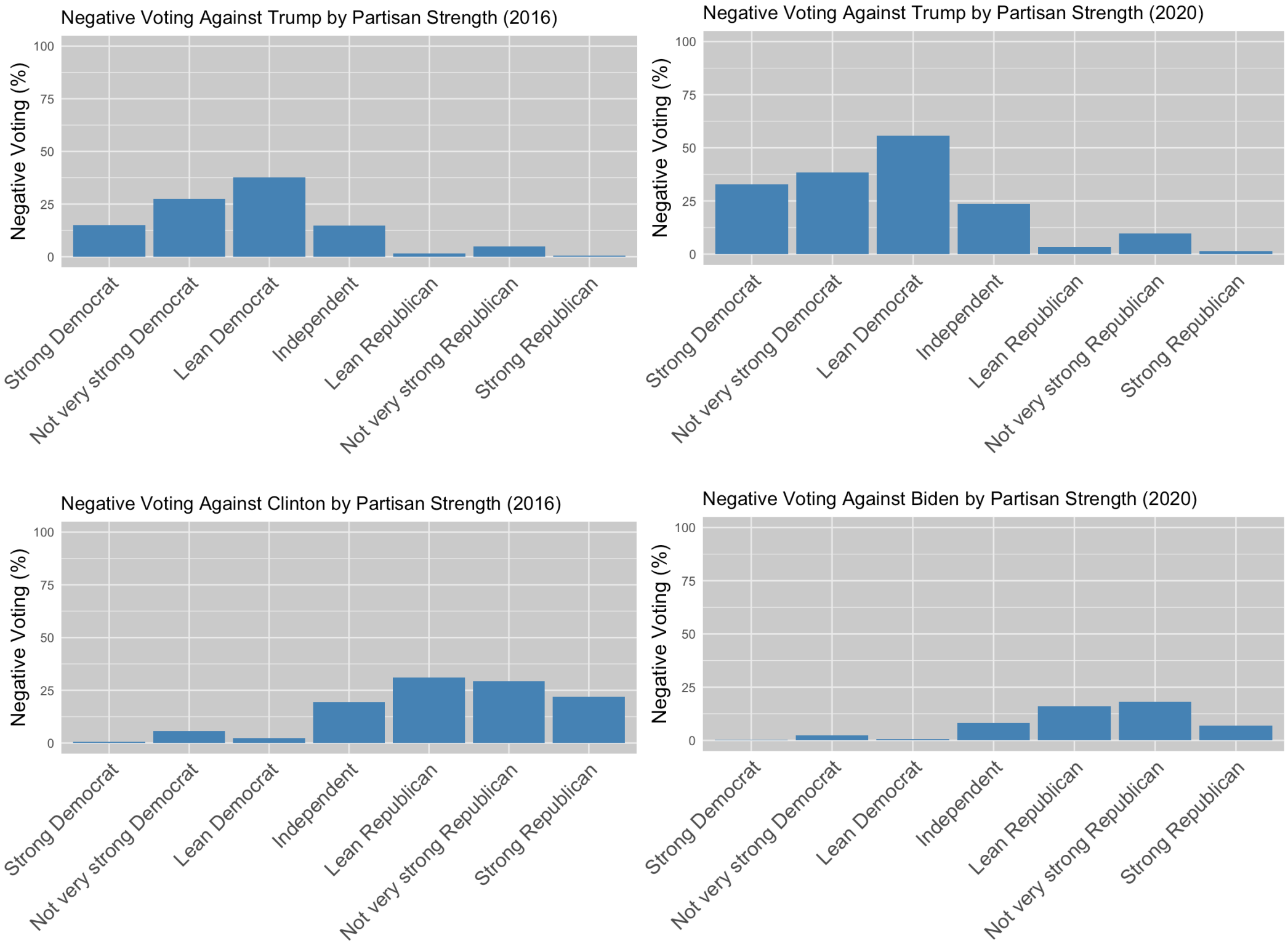

Figure A4. Partisan-Based Favor Voting in 2016 and 2020

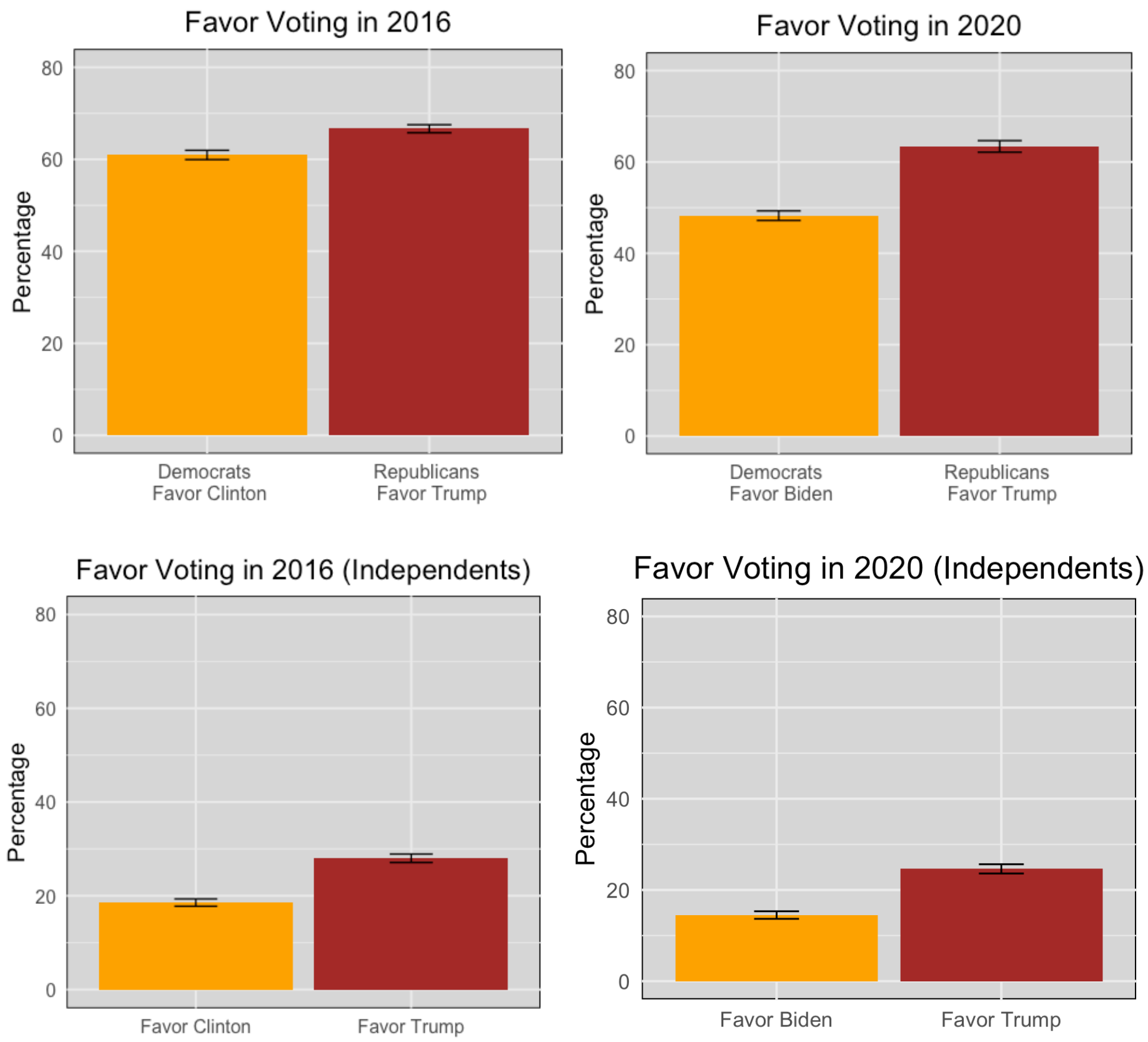

In contrast, as Figure 5 shows, in-group favorability for preferred candidates in 2016 and 2020 among Democrats, Republicans, and Independents was relatively comparable. These plots indicate that across both elections and partisan groups, favorability toward Trump was slightly higher than toward Clinton in 2016 and Biden in 2020. This suggests that many people did not enthusiastically or genuinely support Biden but were instead motivated to vote for him due to their strong disfavor of Trump.

Table A1. Summary Statistics of Anti-Trump Items

| | N | Mean | Std. dev. | Min | Max |
|---|---|---|---|---|---|
| Economy | 4,764 | 2.64 | 1.34 | 1 | 4 |
| Race relations | 4,729 | 2.92 | 1.32 | 1 | 4 |
| Covid-19 | 4,816 | 3.01 | 1.26 | 1 | 4 |
| Ovarall approval | 4,838 | 2.92 | 1.30 | 1 | 4 |

Table A2. Correlation Matrix

| | Economy | Race relation | Covid-19 | Approval |
|---|---|---|---|---|
| Economy | 1 | | | |
| Race relation | 0.91 | 1 | | |
| Covid-19 | 0.88 | 0.94 | 1 | |
| Approval | 0.90 | 0.96 | 0.95 | 1 |

The Alpha scale reliability coefficient is 0.98

Table A3. Anti-Trump Sentiment and Negative Voting Against Trump in 2020

| | (1) | (2) | (3) | (4) | (5) | (6) |
|---|---|---|---|---|---|---|
| | Democrat | | Republican | | Independent | |
| Anti-Trump Sentiment | 0.789*** | | 2.732*** | | 1.791*** | |
| | (0.153) | | (0.314) | | (0.144) | |
| Negative Partisanship | | 0.000 | | 0.040*** | | 0.025*** |
| | | (0.002) | | (0.005) | | (0.002) |
| Perceived Economic Trend | 0.175* | 0.316*** | 0.0568 | 0.259 | 0.271** | 0.543*** |
| | (0.097) | (0.095) | (0.343) | (0.275) | (0.136) | (0.124) |
| **Education (Reference: HS or less)** | | | | | | |
| Some college | 0.008 | -0.019 | 1.980*** | 1.538*** | 0.198 | 0.105 |
| | (0.145) | (0.144) | (0.685) | (0.596) | (0.213) | (0.207) |
| Bachelor or more | 0.076 | 0.0458 | 1.698*** | 1.561*** | 0.410* | 0.478** |
| | (0.149) | (0.149) | (0.650) | (0.580) | (0.221) | (0.215) |
| **Race/Ethnicity (Reference: Black)** | | | | | | |
| White | 0.376** | 0.299* | 2.534** | 2.544* | 0.170 | -0.020 |
| | (0.156) | (0.155) | (1.267) | (1.447) | (0.266) | (0.272) |
| Hispanic | -0.002 | -0.109 | 3.529** | 3.065* | -0.765** | -0.885** |
| | (0.205) | (0.204) | (1.461) | (1.584) | (0.359) | (0.367) |
| Asian | 0.175 | 0.083 | 3.105** | 3.421** | 0.303 | 0.245 |
| | (0.317) | (0.318) | (1.398) | (1.567) | (0.474) | (0.469) |
| **Age (Reference: 51-60)** | | | | | | |
| <30 | 0.832*** | 0.814*** | -0.967 | -0.651 | 0.396 | 0.731*** |
| | (0.182) | (0.183) | (0.966) | (0.812) | (0.269) | (0.261) |
| 31-40 | 0.253 | 0.205 | -1.328* | -0.247 | 0.494* | 0.654** |
| | (0.188) | (0.187) | (0.793) | (0.660) | (0.279) | (0.266) |
| 41-50 | -0.0548 | -0.027 | 0.194 | 0.147 | 0.513* | 0.528** |
| | (0.179) | (0.180) | (0.760) | (0.630) | (0.266) | (0.249) |
| 61-70 | -0.411*** | -0.395** | 0.463 | 0.273 | 0.078 | 0.014 |
| | (0.158) | (0.158) | (0.686) | (0.551) | (0.228) | (0.220) |
| 71+ | -0.197 | -0.216 | -0.395 | 0.189 | -0.149 | -0.205 |
| | (0.184) | (0.184) | (0.782) | (0.705) | (0.296) | (0.292) |
| **Income** | | | | | | |
| Low (<40k) | 0.088 | 0.0356 | -0.600 | 0.670 | -0.367 | -0.130 |
| | (0.194) | (0.194) | (1.497) | (1.138) | (0.276) | (0.267) |
| Medium (40-99.99k) | 0.281 | 0.232 | 0.619 | 0.738 | 0.101 | 0.320 |
| | (0.185) | (0.184) | (1.440) | (1.069) | (0.259) | (0.249) |
| High (>=100k) | 0.512** | 0.473** | 0.109 | 0.658 | 0.381 | 0.591** |
| | (0.203) | (0.202) | (1.451) | (1.085) | (0.294) | (0.280) |
| Born Again Evangelical | -0.315* | -0.465*** | -0.337 | -0.327 | 0.018 | -0.180 |
| | (0.162) | (0.160) | (0.543) | (0.452) | (0.215) | (0.201) |
| News Interest | 0.138 | 0.281 | 0.315 | 0.513 | 0.048 | -0.040 |
| | (0.173) | (0.173) | (0.720) | (0.614) | (0.229) | (0.222) |
| Married | -0.099 | -0.138 | -0.362 | -0.135 | 0.115 | 0.128 |
| | (0.122) | (0.121) | (0.502) | (0.410) | (0.175) | (0.170) |
| Female | -0.176 | -0.139 | 0.244 | 0.242 | -0.060 | 0.0128 |
| | (0.108) | (0.108) | (0.489) | (0.389) | (0.157) | (0.151) |
| Home Ownership | -0.077 | -0.114 | 0.463 | 0.494 | 0.127 | 0.039 |
| | (0.121) | (0.121) | (0.629) | (0.567) | (0.191) | (0.183) |
| Intercept | -4.580*** | -1.871*** | -14.52*** | -7.565*** | -8.208*** | -3.327*** |
| | (0.671) | (0.399) | (2.749) | (2.197) | (0.754) | (0.563) |
| N | 1,778 | 1,757 | 1,073 | 1,069 | 1,301 | 1,279 |

Standard errors in parentheses, *** p<0.01, ** p<0.05, * p<0.1

## Anti-Trump Sentiment Items

1. Approve Donald Trump as president [trumpapp_2020Nov]

"Do you approve or disapprove of the way Donald Trump is handling his job as President?"

2. Approve Donald Trump on race relations [trumpapp_race_2020Nov]

"Do you approve or disapprove of Donald Trump's handling of race relations?"

3. Approve Donald Trump on economy [trumpapp_econ_2020Nov]

"Do you approve or disapprove of Donald Trump's handling of the economy?"

4. Approve Donald Trump on COVID outbreak [trumpapp_covid_2020Nov]

"Do you approve or disapprove of Donald Trump's handling of the coronavirus outbreak?"

The response options are presented on a 4-point Likert scale: Strongly Approve, Somewhat Approve, Somewhat Disapprove, and Strongly Disapprove.

Figure A5. Effect of Age on Negative Voting

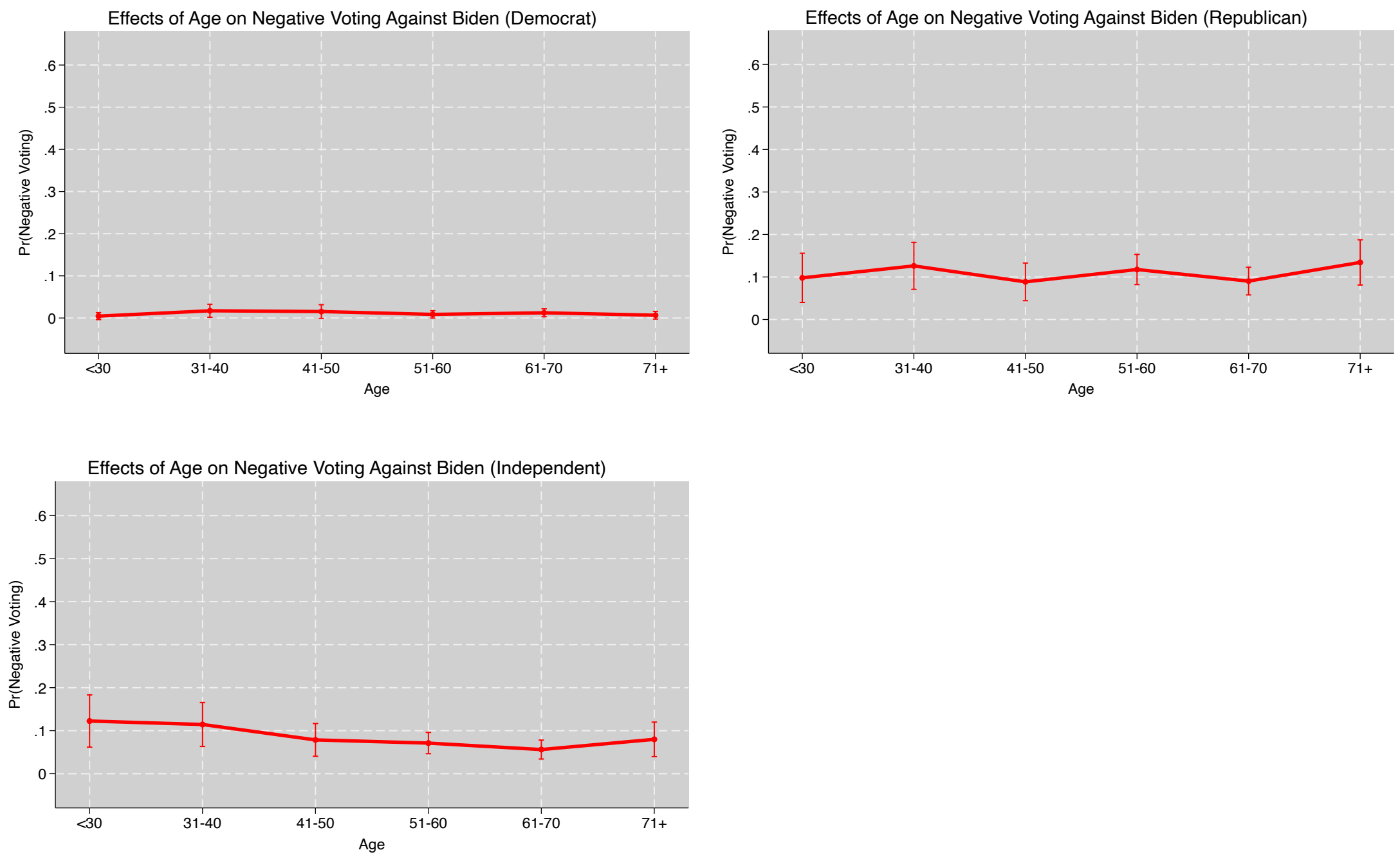

# Figure A6. Effect of Trump Sentiment on Negative Voting

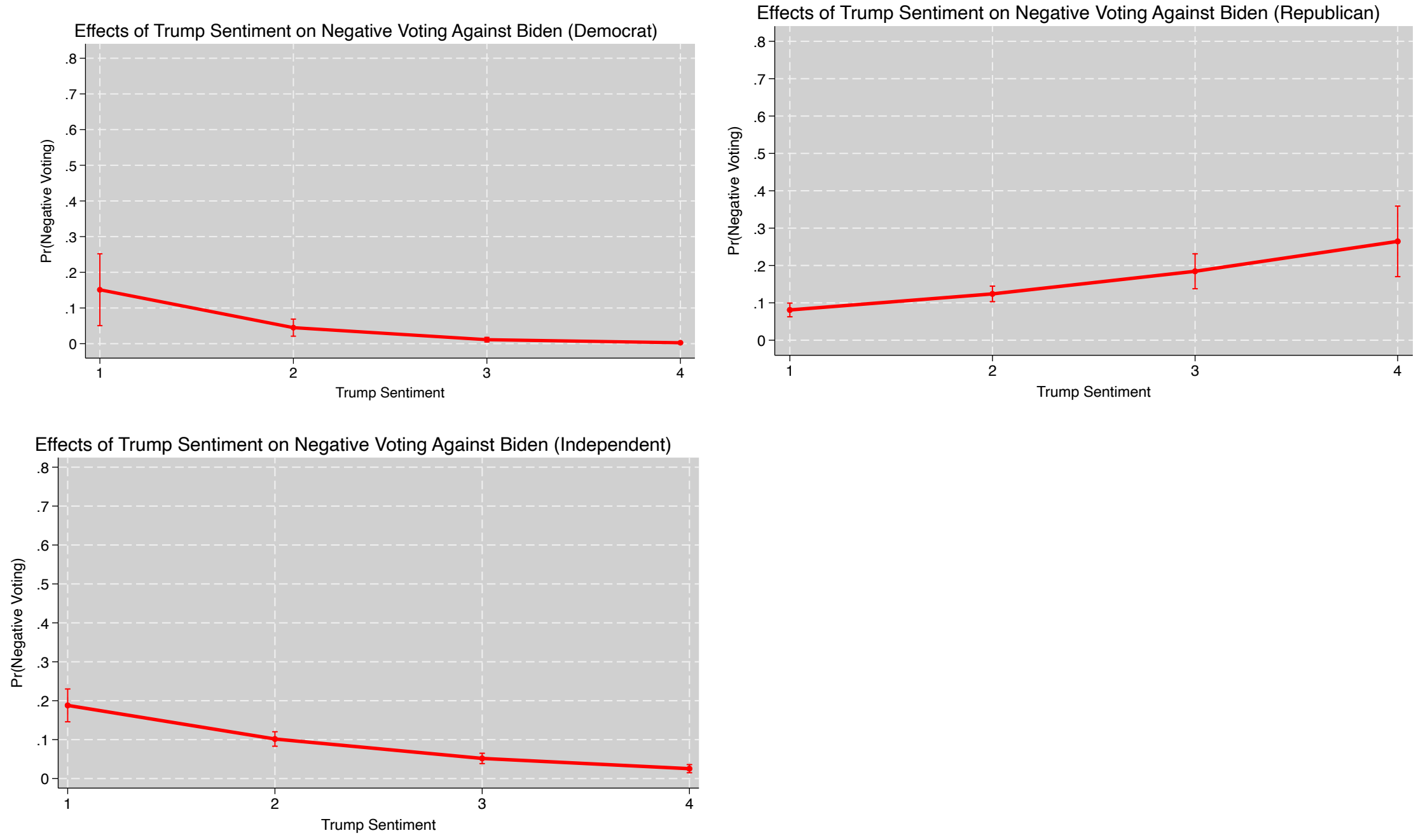